# Custom Chipset and Compact Module Design for a 75-110 GHz Laboratory Signal Source

*Matthew A. Morgan, Tod A. Boyd, and Jason J. Castro*

**Abstract**

We report on the development and characterization of a compact, full-waveguide bandwidth (WR-10) signal source for general-purpose testing of mm-wave components. The MMIC-based multichip module is designed for compactness and ease-of-use, especially in size-constrained test sets such as a wafer probe station. It takes as input a cm-wave CW reference and provides a factor of three frequency multiplication as well as amplification, output power adjustment, and in-situ output power monitoring. It utilizes a number of custom MMIC chips such as a Schottky-diode limiter and a broadband mm-wave detector, both designed explicitly for this module, as well as custom millimeter-wave multipliers and amplifiers reported in previous papers.

## I. Introduction

Commercially-available signal sources (or frequency extenders) for millimeter wave frequencies are often bulky, expensive, and do not provide for direct leveling and/or readout of the signal power. One must usually combine them with external waveguide couplers, power meters, and controllable attenuators in order to set and maintain a desired power level, even approximately. The end result is a mass of waveguide plumbing that is time-consuming to set up and difficult to integrate with test sets designed for small DUTs – such as a wafer probe station.

In this paper we report on the development of a compact WR-10 signal source, comprising an active multiplier chain with an integrated manual attenuator and digital readout displaying the high-frequency output power in real time. The module is optimized for both compactness and ease-of-use in a component-level development environment. As such, it works as a standalone unit, requiring only DC power and a CW reference signal which can be provided at low power by coaxial cable from a standard laboratory synthesizer. No external computer control or software is needed, nor does the low-frequency reference source need to be located close to the device under test.

This work builds upon a previous attempt at such a laboratory signal source [1]. However, that work was completed nearly 15 years ago, and much improvement has been made since then in embedded micro-electronics and millimeter wave amplifier technology. This makes the current re-optimization a timely effort for the next generation of millimeter wave programs.

An overview of the signal path and the module construction is given in Section II. The individual custom MMICs and other components will be described in Sections III and IV. Finally, test data for the complete module will be presented in Section V.

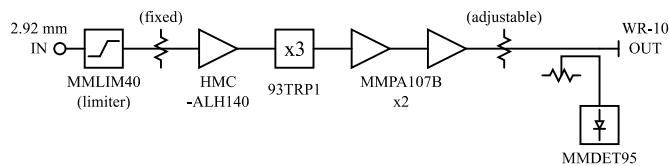

Fig. 1 Block diagram of the multichip module.

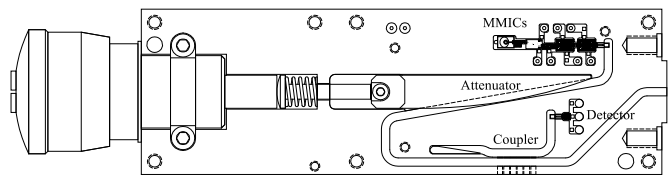

Fig. 2. Physical layout of multichip module.

## II. Module Overview

A block diagram of the signal source module is shown in Fig. 1 and the physical layout of the interior is shown in Fig. 2. The input is provided by a 2.92 mm coaxial connector interface which supports frequencies up to 40 GHz. It is coupled to the MMIC chain through a perpendicular coax to microstrip transition [2].

The first active component is a Schottky-diode MMIC limiter which, along with a fixed attenuator pad, protects the following amplifier stage from being over-driven. The amplifier – commercially available HMC-ALH140 from Hittite Microwave (now Analog devices) – was selected for its moderate output power and flat gain in the band of interest (25-37 GHz), however the desired operating point of about 1 mW input is very close to the absolute maximum input power rating of 4 mW [3]. The limiter was therefore deemed a necessary component to prevent damage to the amplifier during casual laboratory use.

Following the amplifier is a broadband tripler using an anti-parallel diode pair to suppress even harmonics. Reported previously [4]-[5], the tripler is capable of producing up to about 1 mW of power in the 75-110 GHz frequency range when fully driven with >50 mW. In the system reported here, the drive power is somewhat less, about 10 mW, resulting in higher conversion loss but also greater durability.

The tripler output is then amplified with a pair of short gate-length pHEMT GaAs power amplifiers [6]. With a flat gain of 12 dB each, these deliver a maximum output power of around 20 mW across the waveguide bandwidth. This is sufficient to serve as either an RF test tone or a local oscillator to pump a millimeter wave mixer in most laboratory experiments.

Next, a variable attenuator is used to adjust the output power of the signal source as needed for specific test requirements. The attenuator is made using a 100 Ω/□ tapered resistive vane that slides into the waveguide through a longitudinal slot in the broad wall (where the lateral current crossing is zero). The position of the vane is controlled by a manual micrometer.

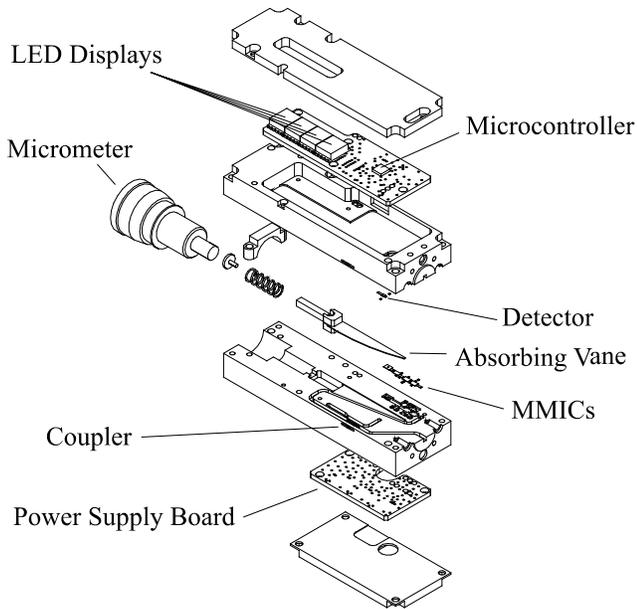

Fig. 3. Exploded view of multichip module.

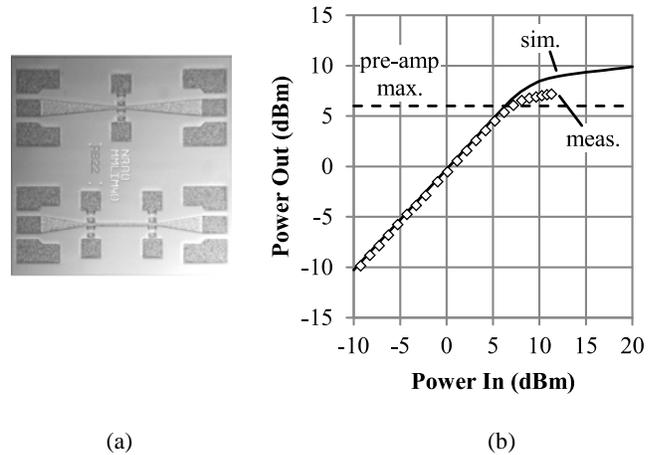

Fig. 4. a) Photograph of microwave diode limiter. Chip size is 1.0 x 1.0 x 0.1 mm. b) Plot of simulated (solid line) and measured (markers) output power vs. input power for the two diode-pair channel at 25 GHz. The horizontal dashed line is the pre-amp max rating. A fixed pad is needed to keep the output below this line.

Following the attenuator is a waveguide coupler based on a multi-aperture Bethe-hole design [7]. The coupled signal is directed to a MMIC square-law detector. The detected voltage passes through the split block with a coaxial feedthrough to a circuit board where it is monitored by an Atmel ATmega328P microcontroller (see Fig. 3). An internal lookup table is used to convert this voltage to power units in dBm, which is then written out to an array of 7-segment LED displays, viewable through a clear optical-grade epoxy window in the lid of the housing.

### III. Custom Chips

A number of custom MMIC chips were required to complete this module. Two of them – the tripler (listed in Fig. 1 as 93TRP1) and post-amplifier (MMPA107B) – were reported in previous publications [4]-[6]. The others, a microwave limiter and a millimeter wave square-law detector, were designed explicitly for this module and were fabricated on a multi-project wafer in the BES Schottky diode MMIC process by United Monolithic Semiconductors (UMS). The design and test results for these two chips are reported below.

*A. Microwave Limiter*

As described in Section II, the microwave pre-amplifier in this module is intended to operate relatively close to its absolute maximum power rating, necessitating a limiter in front of it for protection. A photograph of the chip is shown in Fig. 4a. The design of this limiter is straightforward, comprising a microstrip though-line with anti-parallel diode pairs connected in shunt. Two versions were designed, having either a single pair or two pairs of anti-parallel diodes spaced a quarter-wavelength apart. Both sets were printed on the same chip, allowing the module designer to select between two different levels of limiting simply by rotating the substrate 180 degrees. For this module, the two diode-pair channel is used. In both cases, a microstrip taper is used to better match the impedance of the parallel diodes to 50 Ω.

For simplicity, the limiter is designed for passive operation only – that is, no bias is provided to the diodes. This avoids complex bias tees and makes it easy to ensure broadband operation up to 40 GHz, however the saturation level depends solely on the diode parameters and cannot easily be controlled. Intrinsically, it limits at too high a power level for the pre-amp it is meant to protect. Therefore, the limiter is used in combination with a following attenuator pad to keep the output power delivered to the pre-amp within specs.

Measured vs. simulated performance at 25 GHz input is plotted in Fig. 4b (curves at other frequencies in the operating band are similar). The saturation point of the limiter is about 2 dB lower than was predicted – ironically, an improvement in this case, since higher saturation would have required a larger attenuator and resulted in less overall module gain. The error is likely due to the diode model, which was intended for mixer applications and may be inaccurate for this kind of usage.

Nevertheless, the saturation point of the limiter is still higher than the maximum rating of the pre-amp, so the value of the intervening fixed attenuator was simply adjusted to put the signal in the proper range while still providing an adequate safety margin.

*B. Millimeter Wave Square-Law Detector*

In order to provide a readout of the achieved output power from the module, an RF detector operating from 75-110 GHz is needed. None were found in commercially available chip catalogs, so a custom MMIC was designed to fulfill this requirement also. A schematic and photograph of the chip are shown in Fig. 5.

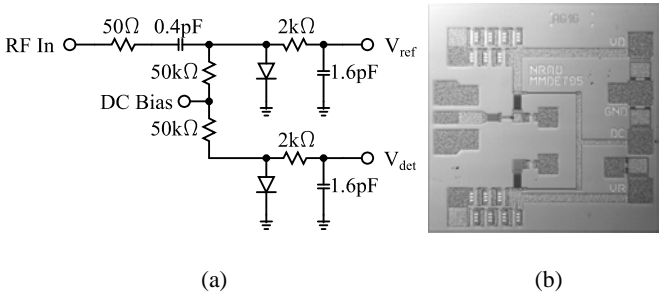

(a)                    (b)

Fig. 5. a) Schematic of the custom millimeter wave detector used in this module. Static protection diodes not shown. b) Photograph of the chip. Size is 1.0 x 1.0 x 0.1 mm.

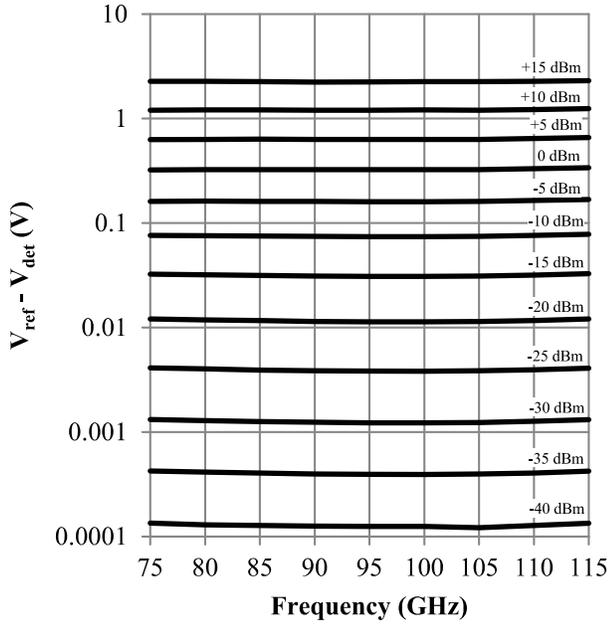

Fig. 6. Predicted performance of millimeter wave detector.

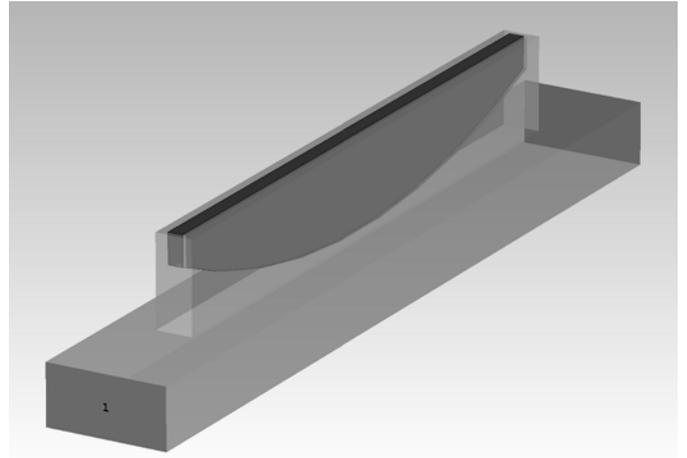

(a)

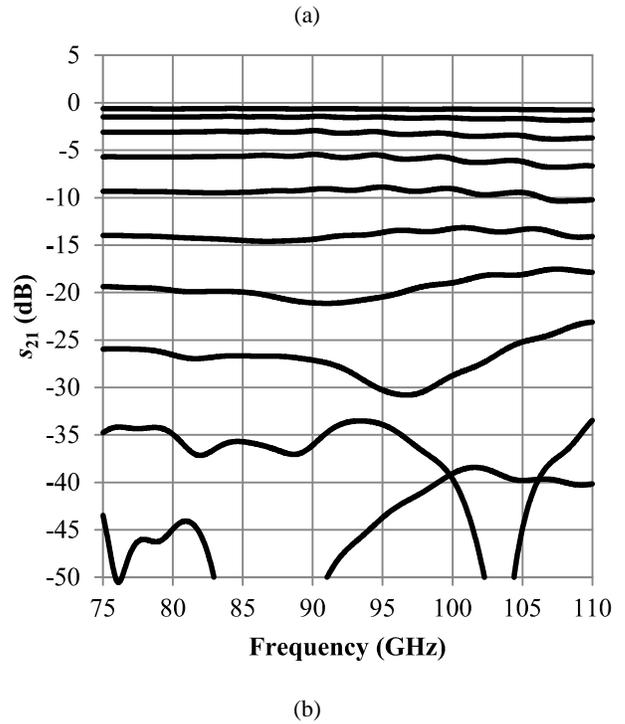

(b)

Fig. 7. a) 3D simulation model of absorbing vane attenuator (shown fully retracted). b) Predicted attenuation in ten equal steps covering the full height of the waveguide.

There are two independent detector diodes, one of which has no input except bias and is used as a reference to correct for internal offsets against which very small detected power levels can be measured. Care was taken during the layout to ensure that both channels are as close to one another as possible.

A 50 Ω resistor on the RF input helps to match the relatively high impedance of the detector diode over a broad frequency range. A 0.4 pF coupling capacitor sets a lower-frequency limit well below the millimeter wave band while acting as a DC block. A small bias current is supplied, to both the primary signal and reference channels, through independent 50 kΩ resistors. Following the detector diodes in each channel is a 2 kΩ resistor and bypass cap to isolate the readout electronics from the detector.

The readout electronics comprise a pair of op-amps designed to generate a large output voltage proportional to the difference between the primary and reference channels of the detector ($V_{det}$ and $V_{ref}$, respectively). This voltage is then monitored by the microcontroller, and converted to a power measurement using a calibrated internal lookup table.

The frequency-independence of this detector is critical as, like any power meter, the module will have no internal knowledge of the frequency being measured. The matching network was therefore designed to provide as near to constant responsivity across the band as possible. The predicted responsivity as a function of frequency is shown in Fig. 6.

The measured performance of the detector chip forms part of the WR-10 source module calibration and will be discussed in Section V.

## IV. Waveguide Components

Following the final, millimeter wave power amplifiers, the output signal is coupled to WR-10 waveguide using an E-field longitudinal probe. The remainder of the primary signal path is in waveguide, and comprises two additional components for output power monitoring and control.

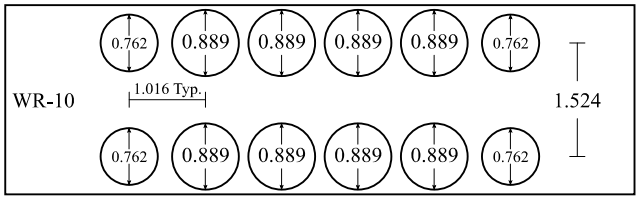

(a)

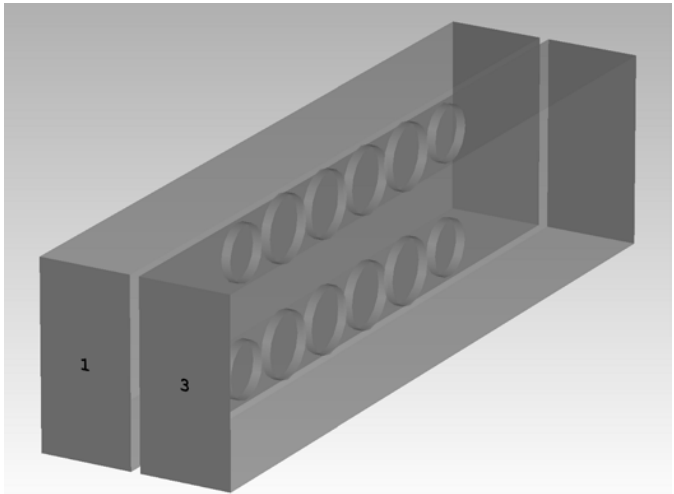

(b)

Fig. 8. a) Physical dimensions (in mm) of Bethe-hole coupler. Wall thickness is 0.127 mm. b) 3D simulation model.

*A. Variable Attenuator*

The first such component is a variable attenuator. It consists of a retractable absorbing vane that enters the waveguide through a slot in the broad wall. The motion of a vane is controlled by a manual micrometer, oriented such that the axis of motion is oblique to the longitudinal axis of the guide, thereby using the full range of motion of the micrometer (about 6.35 mm) to position the vane within the 1.27 mm projected height of the waveguide.

While solid-state, voltage-variable attenuators in MMIC form are readily available (or could be designed, if needed), they typically require two control voltages that are non-linearly dependent upon one another, necessitating some kind of micro-processor control. This in itself is not difficult to implement with an external computer, and may have even proven more convenient in a production environment where computer-control can be used for scripting a pre-determined array of measurements, however in research testing the setup time associated with automation is often an impediment to rapid adaptation or real-time trouble-shooting. For this reason, a manually-controlled variable attenuator that needs no external computer was deemed more desirable in the intended application scenario for this development.

The curvature and sheet resistance of the absorbing vane were selected based on a three-dimensional electromagnetic simulation, given the constraints of the available resistive materials (especially substrate composition and thickness).

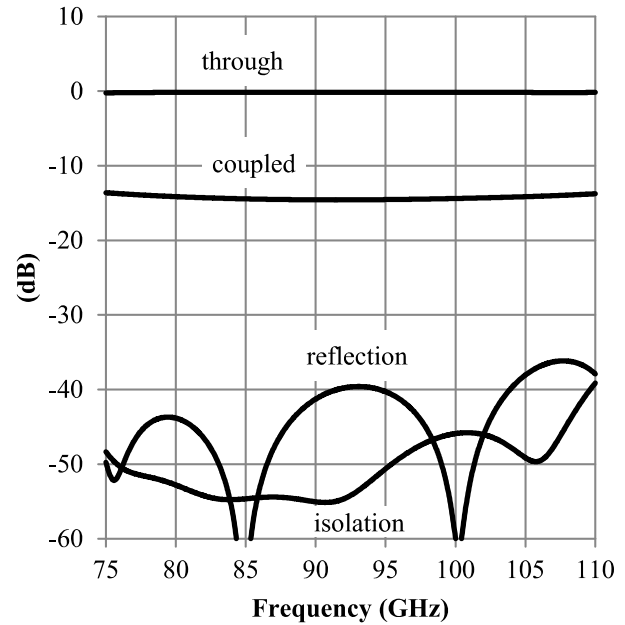

Fig. 9. Predicted performance of Bethe-hole coupler.

The simulation model and predicted performance are shown in Fig. 7. The frequency response becomes somewhat unpredictable at the very deepest levels of attenuation, but is flat for attenuations up to 20 dB.

*B. Beth-Hole Coupler*

A multi-aperture Bethe-hole coupler was designed for this module [7]. The physical dimensions are shown in Fig. 8a. First, a wall thickness of 0.127 mm was selected for manufacturability and achievable coupling for a modest number of apertures. Next, the offset of the apertures from the center-line of the waveguide was optimized to achieve the flattest coupling response per aperture pair across the frequency band. Several pairs of coupling apertures were then arranged 1.016 mm apart, approximately quarter-wavelength spacing, to achieve the desired coupling. Finally, the diameter of the outermost coupling apertures was adjusted to optimize the impedance match. The final simulation model is shown in Fig. 8b.

The predicted performance of the coupler is shown in Fig. 9. It achieves a flat coupling response of about 15 dB, with better than 30 dB directivity over most of the band. Return loss and insertion loss are also near perfect for this component.

To fabricate this part, all twelve coupling apertures were drilled from the outside of the split block through the broad wall of the waveguide. As a consequence, the same holes that collectively provide 15 dB coupling between waveguides internally are present in the external wall of the housing. However, the outside wall of the block is an order of magnitude thicker than the thin internal wall separating the waveguides. Since the apertures are well below cutoff for the frequency band of interest, the coupling through them is exponentially dependent on this thickness. As a result, the leakage through these residual holes in the outer wall which remain as an artifact of the machining process is quite negligible.

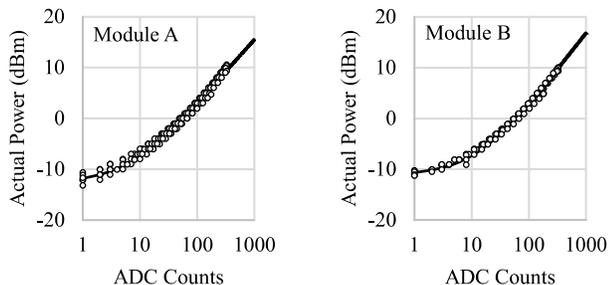

Fig. 10. Measured calibration data and best fit curves for both WR-10 source modules.

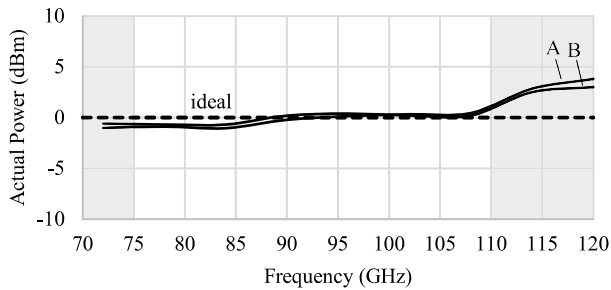

Fig. 11. Actual output power for modules A and B as a function of frequency, given a displayed value of 0 dBm. The shaded areas are outside the waveguide bandwidth.

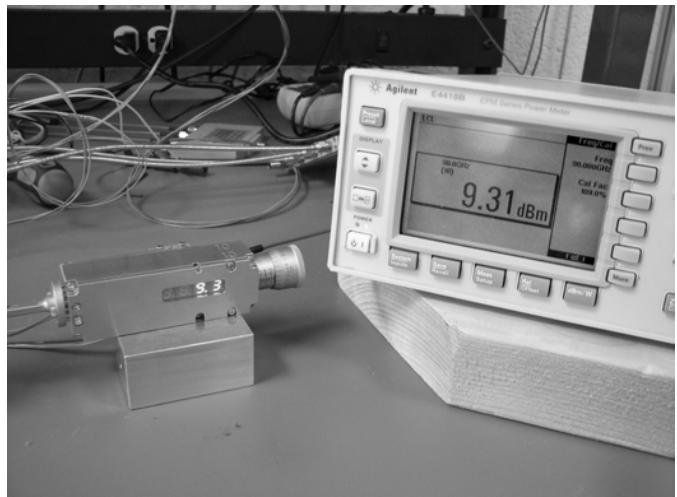

Fig. 12. Photograph of module A delivering a 9.3 dBm CW tone at 90 GHz, with a power meter for comparison.

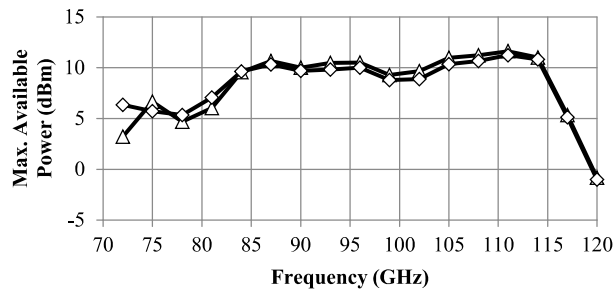

Fig. 13. Maximum measured output power as a function of frequency for module A (triangular markers) and module B (diamonds).

## V. Module Test Data

Two identical modules were built, calibrated and tested. The pair is needed for future experiments with down-converters where both an LO and an RF test signal are required.

Calibration was performed by first programming the ATmega328P microcontroller to display simple ADC counts. The output power as a function of frequency and attenuator micrometer setting was then measured with a power meter and recorded. A logarithmic curve was then fitted to the data and the resulting lookup table programmed into the microcontroller so that the display reads in dBm.

Calibration curves for both modules are shown in Fig. 10. The spread in measured points around a particular power setting result from the combined frequency-dependence of the detector responsivity as well as the coupling curve of the coupler, which were described in Sections III and IV. The overall accuracy of the power monitor is better illustrated by Fig. 11, which shows the actual output power measured with a power meter when the integrated readout displays 0 dBm (1 mW). A photograph of module A in operation illustrates similar agreement between the integrated readout and a calibrated power meter at higher power levels in Fig. 12.

The maximum output power of the modules is determined by the saturation level of the power amplifiers and subsequent losses in the waveguide probe, the attenuator (when the vane is fully retracted), and the coupler. Because the latter two are all-waveguide parts, the probe itself is likely to be the dominant contributor to this loss.

Based on the known saturation level of the amplifiers and estimated losses, we should expect about 10 dBm available power across most the band. The measured maximum output power is shown in Fig. 13 which is in good agreement with this prediction. The roll off below 85 GHz is believed to be due to the power amplifier gain dropping out. Previous measurements in [6] did not show its performance below this frequency.

## VI. Conclusion

A compact WR-10 signal source was designed as a multichip module, and two prototype units were built and tested. Measurements are in good agreement with predictions for both the individual custom MMIC chips and waveguide components as well as the final assemblies. Both modules deliver up to approximately 10 mW of output power in the 75-110 GHz frequency range. The power is manually adjustable and the modules include an integrated power readout to facilitate rapid prototyping and troubleshooting in laboratory measurements of small devices.

To illustrate the utility of this development, a photograph of the two modules being used simultaneously to probe-test a millimeter wave sideband separating mixer is shown in Fig. 14. (Heat sinks have been added in this measurement to better protect the internal electronics from overheating.) Crowding around the probes would have made this measurement setup

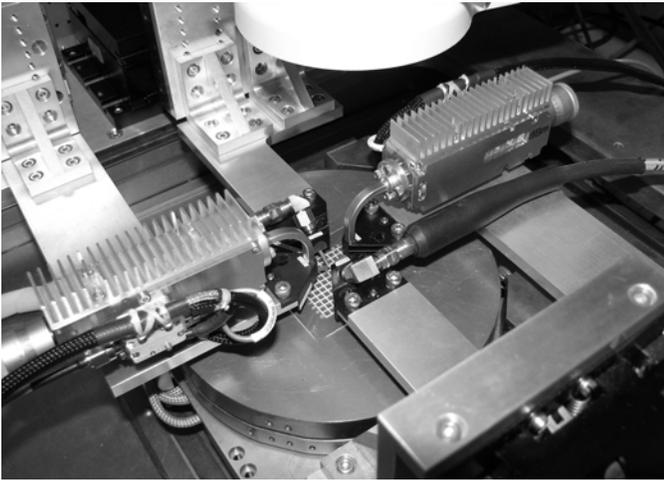

Fig. 14. Photograph of the two multichip modules in use on a probe-station for testing a millimeter wave MMIC mixer.

much more difficult using currently available laboratory sources.

## Acknowledgment

The National Radio Astronomy Observatory is a facility of the National Science Foundation operated under cooperative agreement by Associated Universities Inc.